\documentclass[12pt,a4paper,twoside,prflogo,Research,Versita]{versita}

\newcommand{\asize}{\mathcal{S}}
\newcommand{\SdS}{Schwarzschild--de~Sitter }

\title{Extreme gravitational lensing in vicinity of \SdS black~holes}

\author{Pavel Bakala\inst{1}\email{E-mail: pavel.bakala@fpf.slu.cz}, Petr \v Cerm\' ak \inst{2}, Stanislav Hled\' ik\inst{1}\\Zden\v ek Stuchl\'ik\inst{1}, Kamila Truparov\' a\inst{1}}

\institute{Institute of Physics, Faculty of Philosophy and Science,Silesian University in Opava\\Bezru\v covo n\' am. 13, CZ-746\,01 Opava, Czech Republic\and Institute of Computer Science, Faculty of Philosophy and Science,
Silesian University in Opava\\Bezru\v covo n\' am. 13, CZ-746\,01 Opava, Czech Republic}

\abstract{
We have developed a realistic, fully general relativistic computer code to simulate optical projection in a strong, spherically symmetric gravitational field. The standard theoretical analysis of optical projection for an observer in the vicinity of a Schwarzschild black hole is extended to black hole spacetimes with a repulsive cosmological constant, i.e, \SdS spacetimes. Influence of the cosmological constant is investigated for static observers and observers radially free-falling from the static radius.
Simulations include effects of the gravitational lensing, multiple images, Doppler and gravitational frequency shift, as well as the intensity amplification. The code generates images of the sky for the static observer and a movie simulations of the changing sky for the radially free-falling observer. Techniques of parallel programming are applied to get a high performance and a fast run of the BHC simulation code.}


\PACS{code}

\begin{document}

\firstpage{1}
\maketitle \setcounter{page}{1}%

\section{Introduction}

General relativistic deflection of light and lensing effects in gravitational field of stars were firstly investigated by Einstein \cite{ein01}. In the vicinity of relativistic compact objects (black holes or neutron stars)  these effects have strong influence on properties of the optical projection which become different than those of the optics in the flat spacetime as we experience it in our everyday life \cite{cunn01}. 
Several authors have developed ray-tracing or simulation computer codes for modeling general relativistic optical projection in the vicinity of rotating or non-rotating black holes and neutron stars without presence of a cosmological constant, see, e.g.,  \cite{BEN1,HAM1,KWR,nem01,NRHK,ohan01,VIE1}.

Recent observations indicate the cosmic expansion accelerated by a dark energy that can be described by a repulsive cosmological constant, $\Lambda > 0$ \cite{Kra-Tur:1995:GenRelGrav:,Ost-Ste:1995:Nature:}. We investigate the influence of $\Lambda > 0$ on the appearance of distant universe for observers in close vicinity of nonrotating \SdS (SdS) black holes. In order to obtain a good qualitative picture of the $\Lambda$ influence, our simulations have been performed with unrealistically high values of $\Lambda$.

\section{Schwarzschild--de~Sitter geometry}
The line element of the SdS spacetime has in the standard
Schwarzschild coordinates and geometric units ($c=G=1$) the form
\begin{equation}
\mathrm{d}s^2=-\left(1-\frac{2M}{r}-\frac{\Lambda}{3}r^2\right)\mathrm{d}t^2
 +\left(1-\frac{2M}{r}-\frac{\Lambda}{3}r^2\right)^{-1}\mathrm{d}r^2
+ r^2(\mathrm{d}\theta^2 + \sin^2\theta\,\mathrm{d}\phi^2), \label{metrika}
\end{equation}
where $M$ is mass of the central black hole, and $\Lambda \sim
{10}^{-56}\,\mathrm{cm}^{-2}$ is the repulsive cosmological constant. It is
advantageous to introduce a dimensionless cosmological parameter $y$ by the
relation $y =\textstyle\frac13\Lambda M^2$.
The location of horizons is given by the condition $g_{\mathrm{tt}}=0$.
Two event horizons exist for $y \in (0,y_{\mathrm{crit}})$, where $y_{\mathrm{crit}}=1/27$.
The black hole and the cosmological horizons are located at
\begin{equation}
   r_{\mathrm{h}}=\frac{2}{\sqrt{3y}}\cos{\frac{\pi+\xi}{3}},  \qquad
   r_{\mathrm{c}}=\frac{2}{\sqrt{3y}}\cos{\frac{\pi-\xi}{3}},
\end{equation}
respectively, where
\begin{equation}
    \xi=\cos^{-1}{3\sqrt{3y}}.
\end{equation}
The spacetime is dynamic at $r<r_{\mathrm{h}}$ and $r>r_{\mathrm{c}}$. The static radius, (a hypersurface where the gravitational attraction of the black hole is balanced
by the cosmic repulsion) is located at
\begin{equation}
    r_{\mathrm{s}}=y^{-\frac{1}{3}}.
\end{equation}
With increasing value of $y$, the horizons aproach to each other. In the critical case of
$y=y_{\mathrm{crit}}=1/27$, the horizons and the static radius coincide at $r_{\mathrm{h}}=3$. For of $y>1/27$,
the spacetime is dynamic at $r>0$, and describes a naked singularity
 \cite{SH}. We consider only spacetimes admitting existence of static observers that have $y<1/27$.
\section{Optical projection in \SdS spacetimes}
Construction of relativistic optical projection consists of finding all null geodesics connecting the source and the observer, i.e., solving the so-called emitter-observer problem. An observer will see the image generated by the concrete geodesic in direction tangent to the photon trajectory in observer's local frame, therefore given by space part of locally measured 4-momentum of photons $p^{(\mu)}_{\mathrm{obs}}$. Directional angle $\alpha$ related to the outward radial direction and the frequency shift $g$ of the photon (the ratio of observed and emitted energy) are given by the general relations
\begin{eqnarray}
     \cos \alpha &=& -\frac{p^{(r)}_{\mathrm{obs}}}{p^{(t)}_{\mathrm{obs}}}\,, \qquad  g =  \frac {p^{(t)}_{\mathrm{obs}}}{p^{(t)}_{\mathrm{source}}}\,.
\label{gen_Angle_shift}
\end{eqnarray}
The indeces  `obs' (observer) and `source' denote the components locally measured by an observer or a source.
In the SdS spacetimes null geodesics are characterised by the impact parameter $b$ defined as the ratio of  constants of motion, $b\equiv\Phi/\mathcal{E}$ \cite{SH,SP}. For an observer located at $r_{\mathrm{obs}}$, $\alpha$ and $g$ are functions of $b$ only, as shown in Appendix.

Due to spherical symmetry of the SdS geometry (\ref{metrika}), it is sufficient to consider only sources and observers located in the equatorial plane. Considering observers located at $\phi=0$, $\Delta\phi$ along geodesics connecting the source and the observer reads
\begin{equation}
\Delta\phi=-\phi_{source} - 2k\pi,
\label{MainEQ}
\end{equation}
where $\phi_{source}$ is an angular coordinate of the source. The image order $k$ takes values of $0,1,2,\ldots,+\infty$ for geodesics orbiting the central black hole clockwise, and $-1,-2,\ldots, -\infty$ for geodesics orbiting the central black hole counter-clockwise. The first direct and indirect images correspond to $k=0$ and $k=-1$, respectively.

Photon motion in the SdS spacetimes is governed by the Binet formula \cite{BCHST,SP}
\begin{equation}
\frac{d \phi}{du}=\pm \frac{1}{\sqrt{b^{-2}-u^{2}+2u^{3}+y}},
\label{Binet}
\end{equation}
where $u=r^{-1}$. The critical impact parameter, $b_{\mathrm{c}} =\sqrt{27/\left(1-27y\right)}$, corresponds to the circular photon geodesic, which is located at $r_{ph}=3M$ for arbitrary value of $\Lambda$ \cite{SH,SP}. Photons coming from distant universe with $b<b_{\mathrm{c}}$ end up in the central singularity, while photons with $b>b_{\mathrm{c}}$ return back towards the cosmological horizont \cite{BCHST,SP}.
Using the term under the square root in (\ref{Binet}) as a motion reality condition, a straightforward calculation yields relation for a turning point $r_{\mathrm{t}}$ of geodesics with $b>b_{\mathrm{c}}$ :
\begin{equation}
r_{\mathrm{t}}=\frac{2}{\sqrt{3(y+b^{-2})}}\cos\left[ \frac{1}{3} \arccos \left(-3\,\sqrt{3\,(y+b^{-2})} \right) \right]
\end{equation}
and a relation for the maximum impact parameter $b_{\mathrm{max}}$ for an observer located at given $r_{\mathrm{obs}}$,
\begin{equation}
b_{\mathrm{max}}=\frac{1}{\sqrt{u^{2}_{obs}-2u^{3}_{obs}-y\,\,}}\,.
\label{bmax}
\end{equation}
Therefore, $\Delta\phi$ can be expressed as
\begin{equation}
      \Delta \phi(u_{\mathrm{source}},u_{\mathrm{obs}},b)=\mp\,\int_{u_{\mathrm{source}}}^{u_{\mathrm{obs}}}\frac{du}
      {\sqrt{b^{-2}-u^2+2u^3+y\,\,}}\,\,,
      \label{IBF1}
\end{equation}
for geodesics with $b<b_{\mathrm{c}}$, or for geodesics passing the observer position ahead of the turning point. For geodesics passing observer position beyond the turning point $\Delta\phi$ can be expressed as
\begin{equation}
      \Delta \phi(u_{\mathrm{obs}})=\pm\,\int_{u_{\mathrm{turn}}}^{u_{\mathrm{obs}}}\frac{du}
      {\sqrt{b^{-2}-u^2+2u^3+y\,\,}} \mp \int_{u_{\mathrm{source}}}^{u_{\mathrm{turn}}}\frac{du}  	         {\sqrt{b^{-2}-u^2+2u^3+y\,\,}}\,\,.
      \label{IBF2}
\end{equation}
In (\ref{IBF1}) and (\ref{IBF2}) , the upper (lower) sign corresponds to geodesics orbiting clockwise (counter-clockwise).
These integrals express $\Delta \phi$ along the photon path as a function $F\left(b,u_{obs},u_{source},y\right)$.
Equation (\ref{MainEQ}) can then be rewriten in the following way :
\begin{equation}
F\left(b,u_{obs},u_{source},y\right)+\phi_{source} + k2\pi =0\,\,.
\label{FinalEQ}
\end{equation}
This equation, that can be understood as an integral equation with an eigenvalue $b$  determines $b$ as an implicit function of the source and observer coordinates, the image order and $\Lambda$. Unfortunately,  $F\left(b,u_{obs},u_{source},y\right)$ can be expressed in terms of elliptic integrals only, and do not allow to obtain  an explicit formula for  $b\left(\phi_{source},k,u_{obs},u_{source},y\right)$. Consequently, the simulation code uses standard numerical integration and root finding methods.

\section{Numerical solution and code outputs}

Final equation (\ref{FinalEQ}) was numerically solved using the BHC code for optical projection of observers located in the  static region between the horizons, as well as for observers located in the dynamic region under the black hole horizon. In the static region, static  observers and observers radially free-falling from the static radius have been considered. In the dymamic region under the black hole horizon, where static observers cannot exist, the optical projection was constructed for observers radially free-falling only. We consider $\Lambda=5.10^{-3}$, $\Lambda=10^{-2}$ and pure Schwarzschild case with $\Lambda=0$.

\begin{figure}[h!]
\begin{minipage}[b]{.48\hsize}
\centering
\includegraphics[width=\hsize]{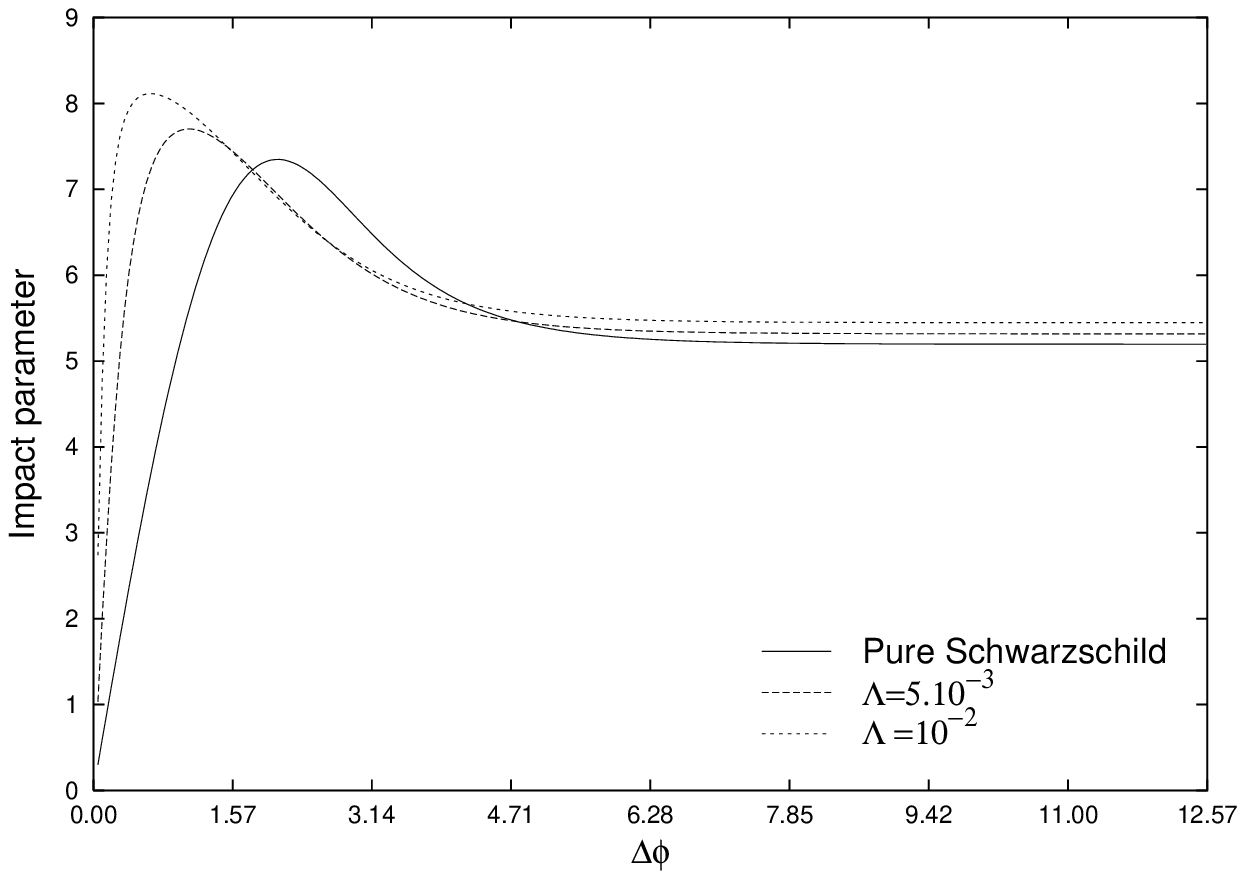}
\end{minipage}\hfill%
\begin{minipage}[b]{.24\hsize}
\centering
\includegraphics[width=\hsize]{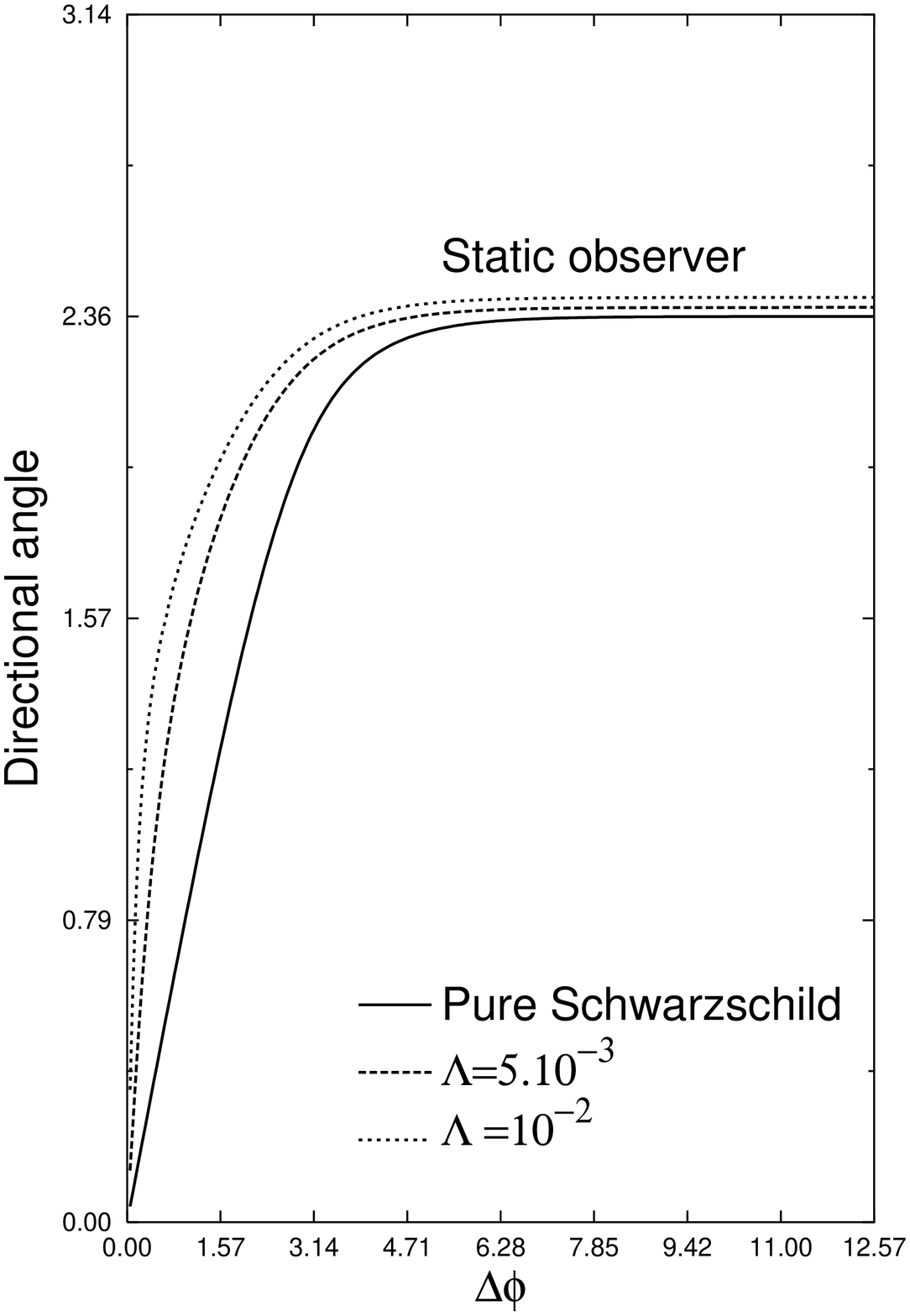}
\end{minipage}\hfill%
\begin{minipage}[b]{.24\hsize}
\centering
\includegraphics[width=\hsize]{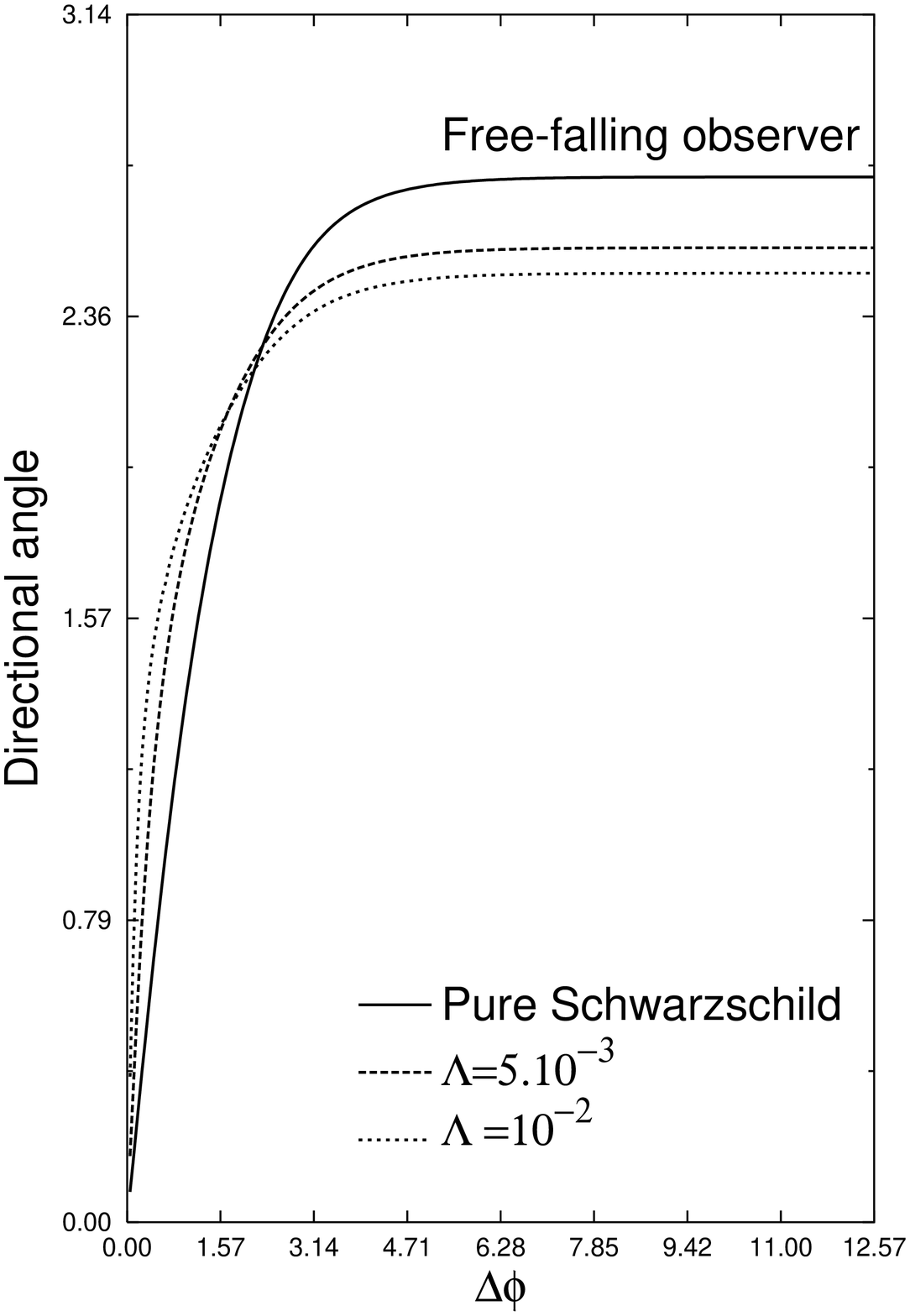}
\end{minipage}
\noindent
\caption{Optical projection for observers located at the static region above the circular photon orbit at $r_{\mathrm{obs}}=6M$.
\textbf{Left panel:} Impact parameter $b$ as a function of $\Delta \phi$. \textbf{Right panels:} Directional angles for static and radially free-falling observers.}

\label{P1}
\end{figure}

\begin{figure}[h!]
\begin{minipage}[b]{.48\hsize}
\centering
\includegraphics[width=\hsize]{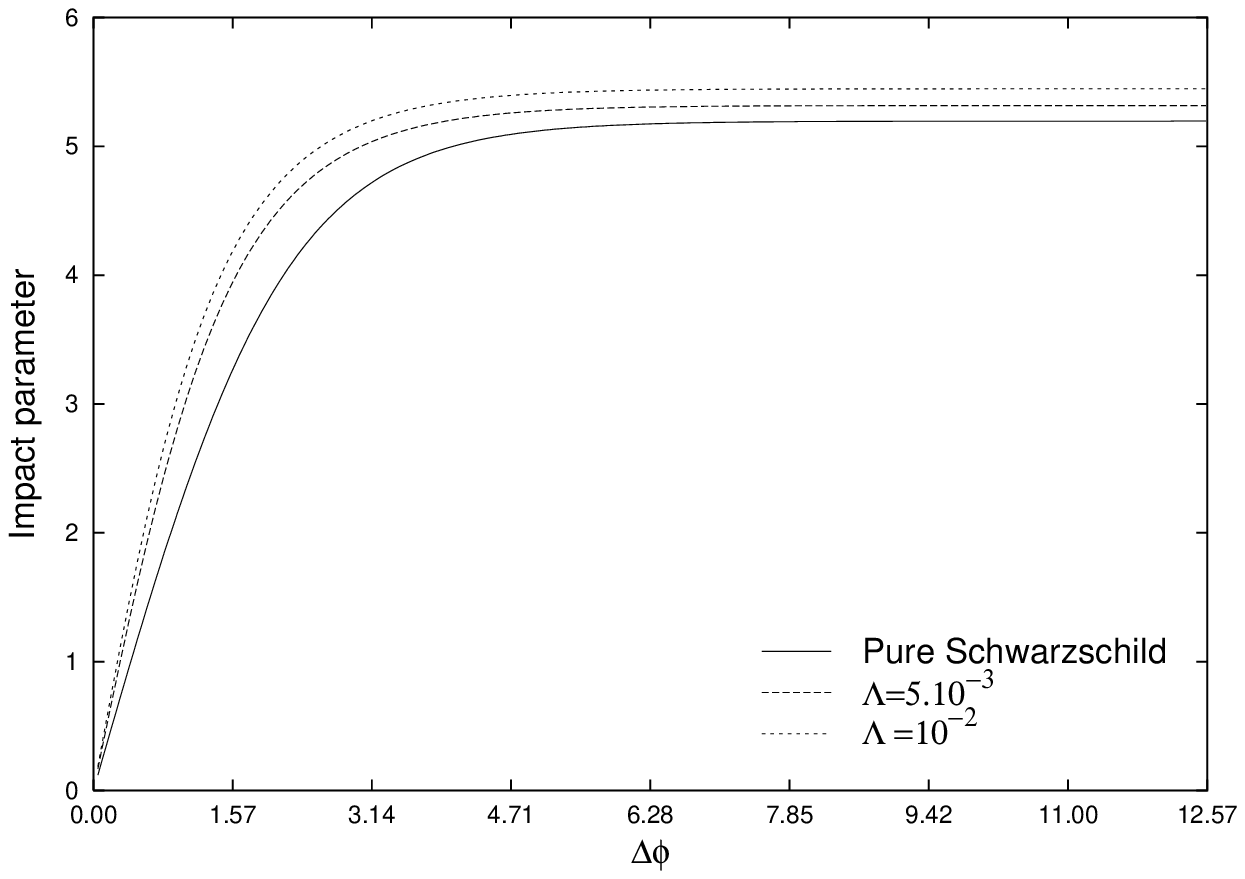}
\end{minipage}\hfill%
\begin{minipage}[b]{.48\hsize}
\centering
\includegraphics[width=\hsize]{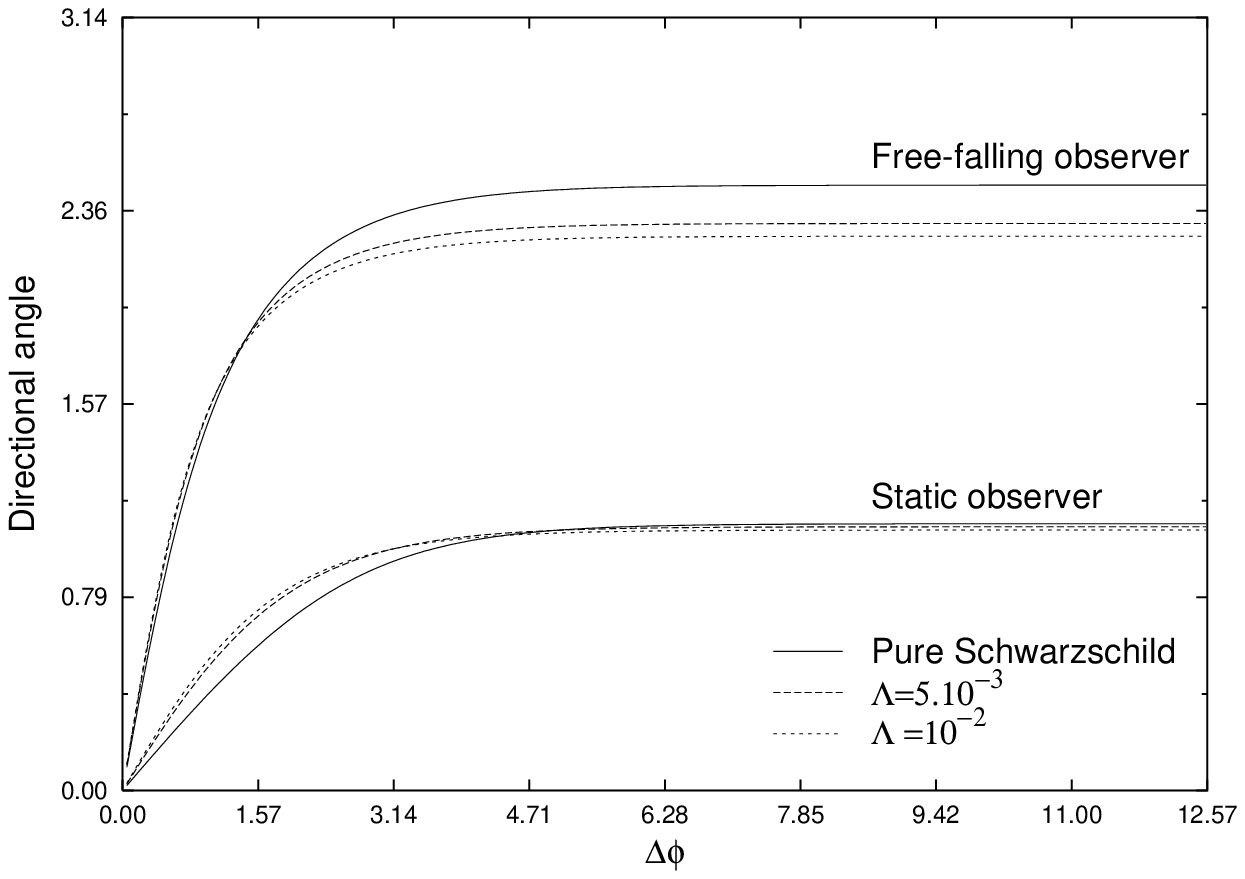}
\end{minipage}
\noindent
\caption{Optical projection for observers located in the static region under the circular photon orbit at $r_{\mathrm{obs}}=2.4M$.
\textbf{Left panel:} Impact parameter $b$ as a function of $\Delta \phi$. \textbf{Right panel:} Directional angles for static and radially free-falling observers.}

\label{P2}
\end{figure}

\begin{figure}[h!]
\begin{minipage}[b]{.48\hsize}
\centering
\includegraphics[width=\hsize]{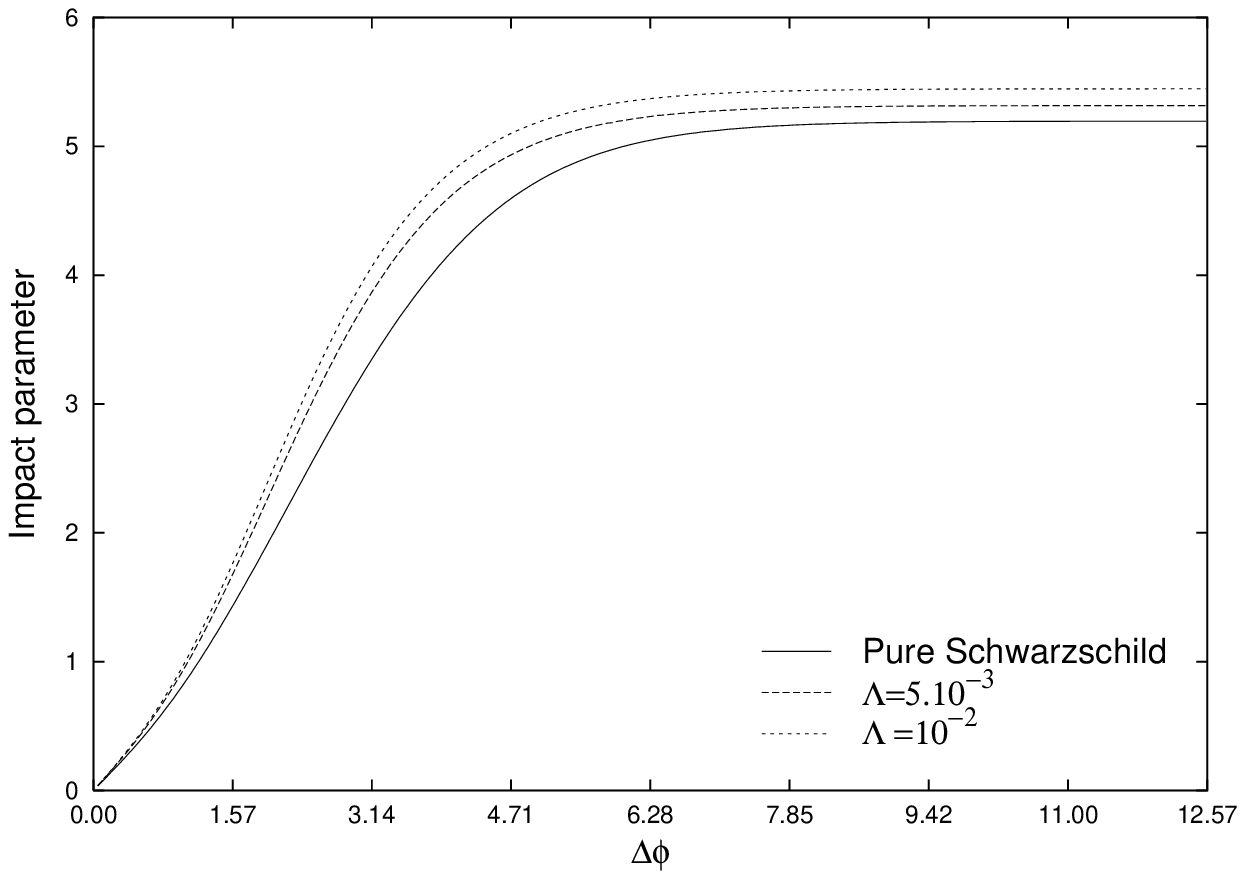}
\end{minipage}\hfill%
\begin{minipage}[b]{.48\hsize}
\centering
\includegraphics[width=\hsize]{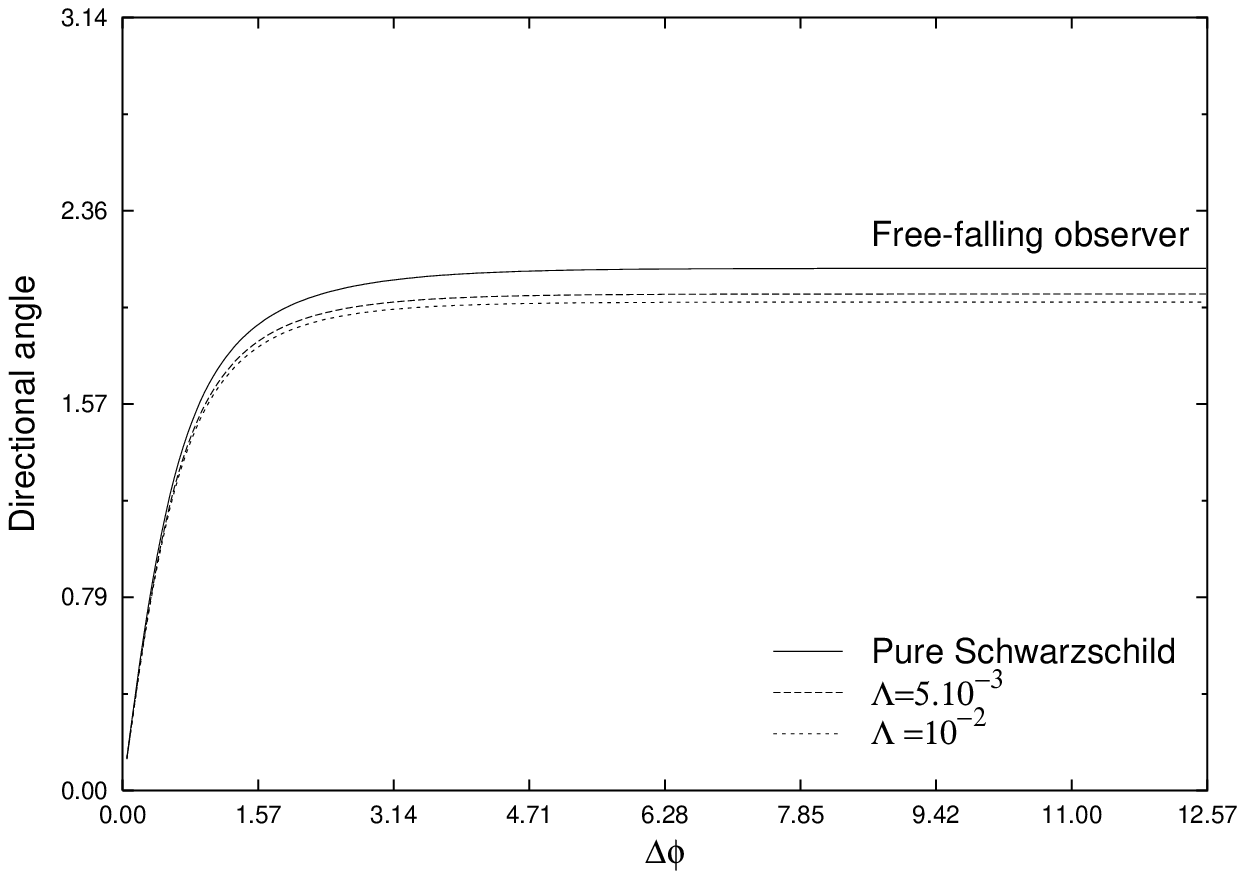}
\end{minipage}
\noindent
\caption{Optical projection for observers located under the black hole horizon at $r_{\mathrm{obs}}=0.7M$.
\textbf{Left panel:} Impact parameter $b$ as a function of $\Delta \phi$. \textbf{Right panels:} Directional angle for  a radially free-falling observer.}

\label{P3}
\end{figure}

Left panels of Figs. \ref{P1}, \ref{P2}, and \ref{P3} show $b$ as a function of $|\Delta \phi|$ along the appropriate geodesic connecting distant source and observer located at $r_{\mathrm{obs}}$. The right panel of these figures show $\alpha$ for static and radially free-falling observers. If we consider spherical symmetry of the problem, equation (\ref{MainEQ}) implies, that $|\Delta \phi|\le\pi$ corresponds to the first direct images of distant sources, whereas larger $|\Delta \phi|>\pi$ corresponds to images with higher order.

As shown in Fig. \ref{P1}, for observers located above the circular photon orbit $b$ increases up to a maximum value $b_{\mathrm{max}}$ given by equation (\ref{bmax}), then decreases, and asymptoticaly aproaches $b_{\mathrm{c}}$ from above.
Values $|\Delta \phi| \le |\Delta\phi(b_{max})|$ correspond to ingoing geodesics with $b<b_{\mathrm{c}}$, or to geodesics passing the observer position ahead the turning point. Values $|\Delta \phi| > |\Delta\phi(b_{max})|$ correspond to geodesics passing the observer position beyond the turning point. Situation is different for observers located under the circular photon orbit where only geodesics with $b<b_{\mathrm{c}}$ can exist. In this case $b$ increases monotonously
and asymptoticaly aproaches to $b_{\mathrm{c}}$ from below.

In all the cases, $\alpha$ monotonously increases up to a maximum value, which determines the black region on the observer's sky. We determine the size of the black region as a function of $r_{\mathrm{obs}}$ and $\Lambda$ in the next section.

\begin{figure}[h!]
\begin{minipage}[b]{.60\hsize}
\centering
\includegraphics[width=\hsize]{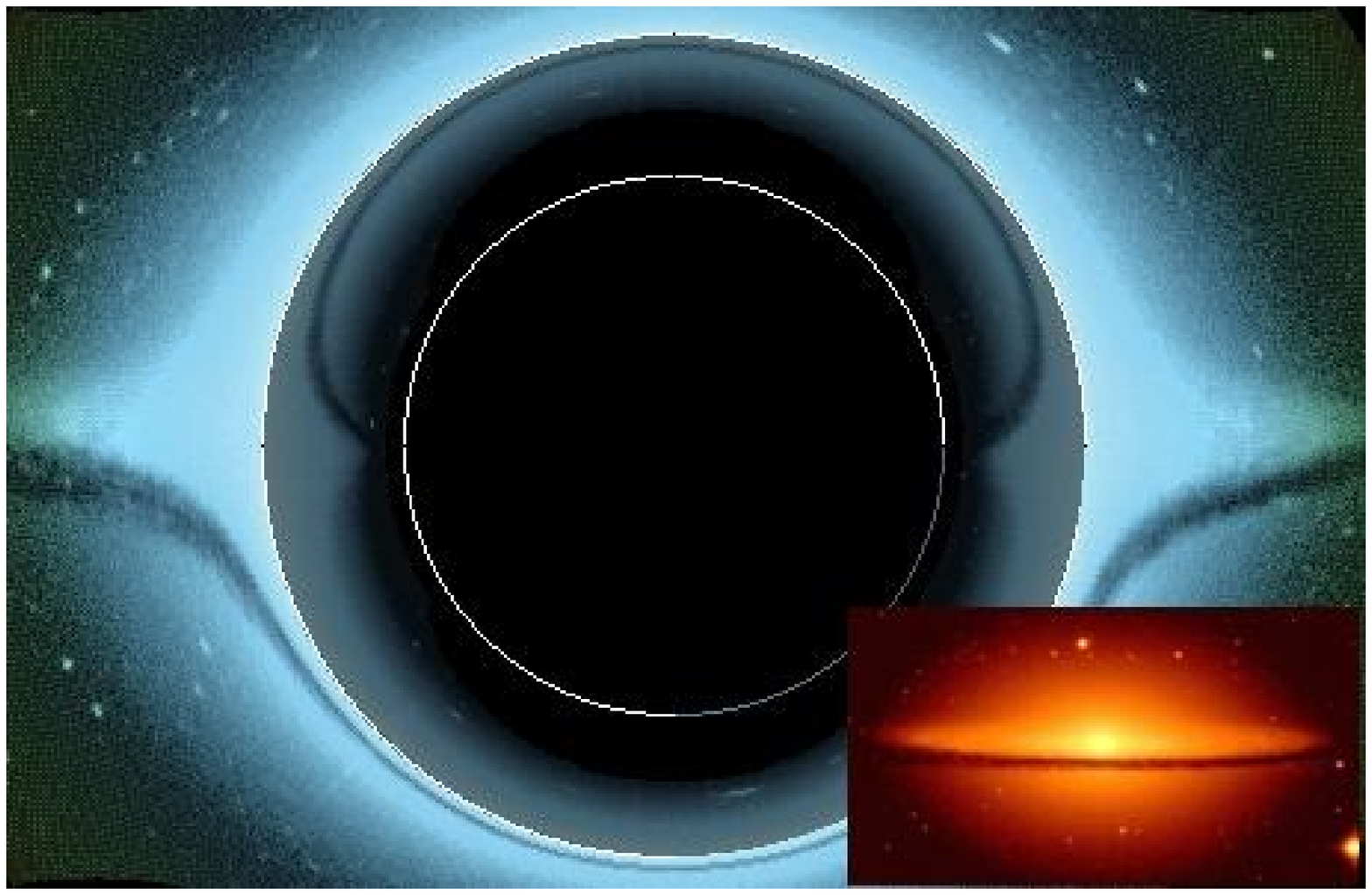}
\end{minipage}\hfill%
\begin{minipage}[b]{.39\hsize}
\centering
\includegraphics[width=\hsize]{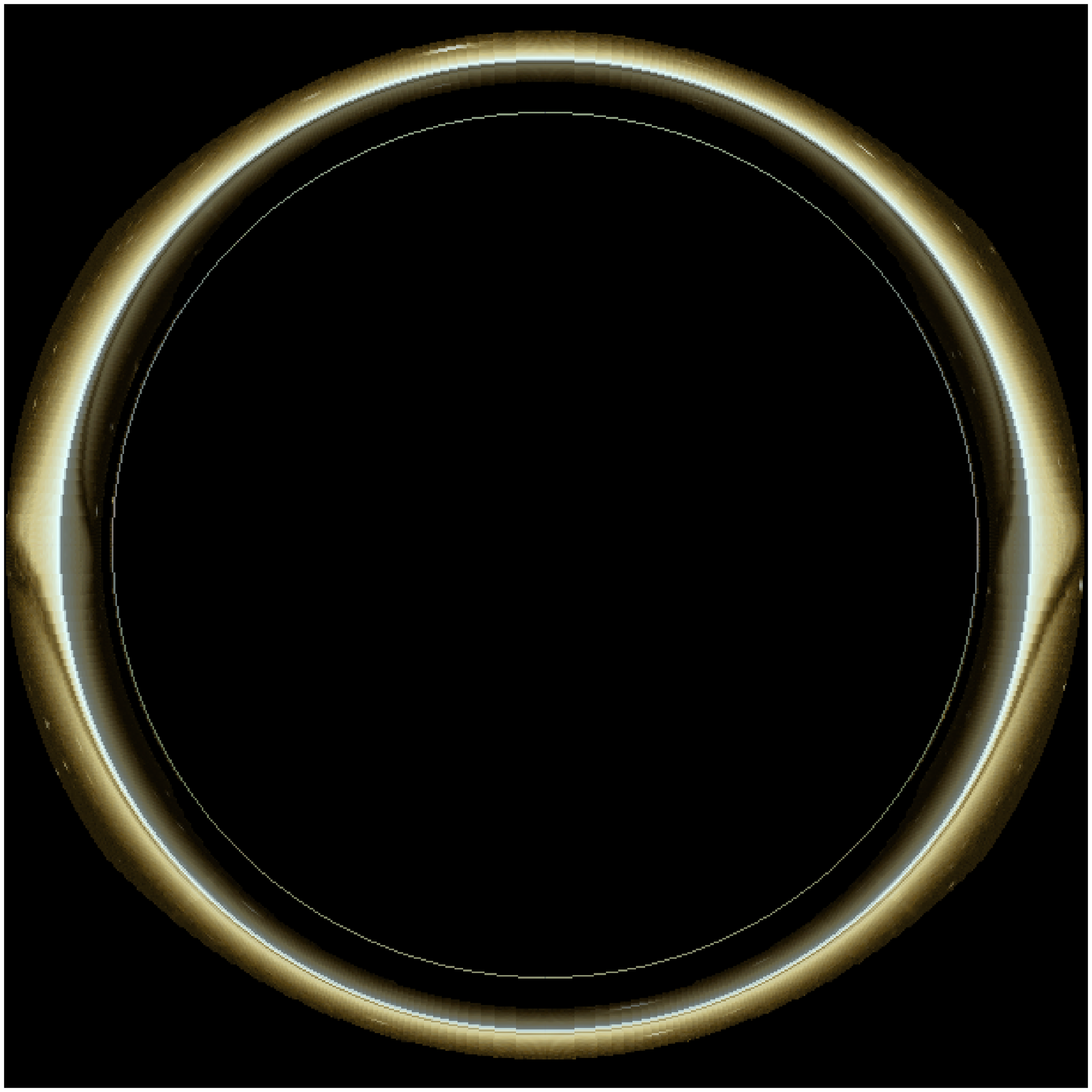}
\end{minipage}
\noindent
\caption{Simulated appearance of M104 ``Sombrero'' located behind the black hole.
\textbf{Left panel:} for a radially free-falling observer at $r_{\mathrm{obs}}=20M$ in a pure Schwarzschild case (with nondistorted image in the right-bottom corner). \textbf{Right panel:} for a static observer at $r_{\mathrm{obs}}=5M$ with $\Lambda=10^{-3}$.}

\label{S1}
\end{figure}

Fig. \ref{S1} shows samples of visualisation outputs of the BHC code, simulated optical projection of a well-known galaxy M104 ``Sombrero'', virtually located behind the black hole on the optical axis. Images show typical lensing effects as the first and second Einstein rings, first direct image, first indirect inverted image and mergence of higher order images with the second Einstein ring around the black region.
Another static images, as well as dynamic simulations, may be downloaded from our web site \cite{our_web}.

\section{Apparent angular size of the black hole}

\begin{figure}[h!]
\begin{minipage}[b]{.48\hsize}
\centering
\includegraphics[width=\hsize]{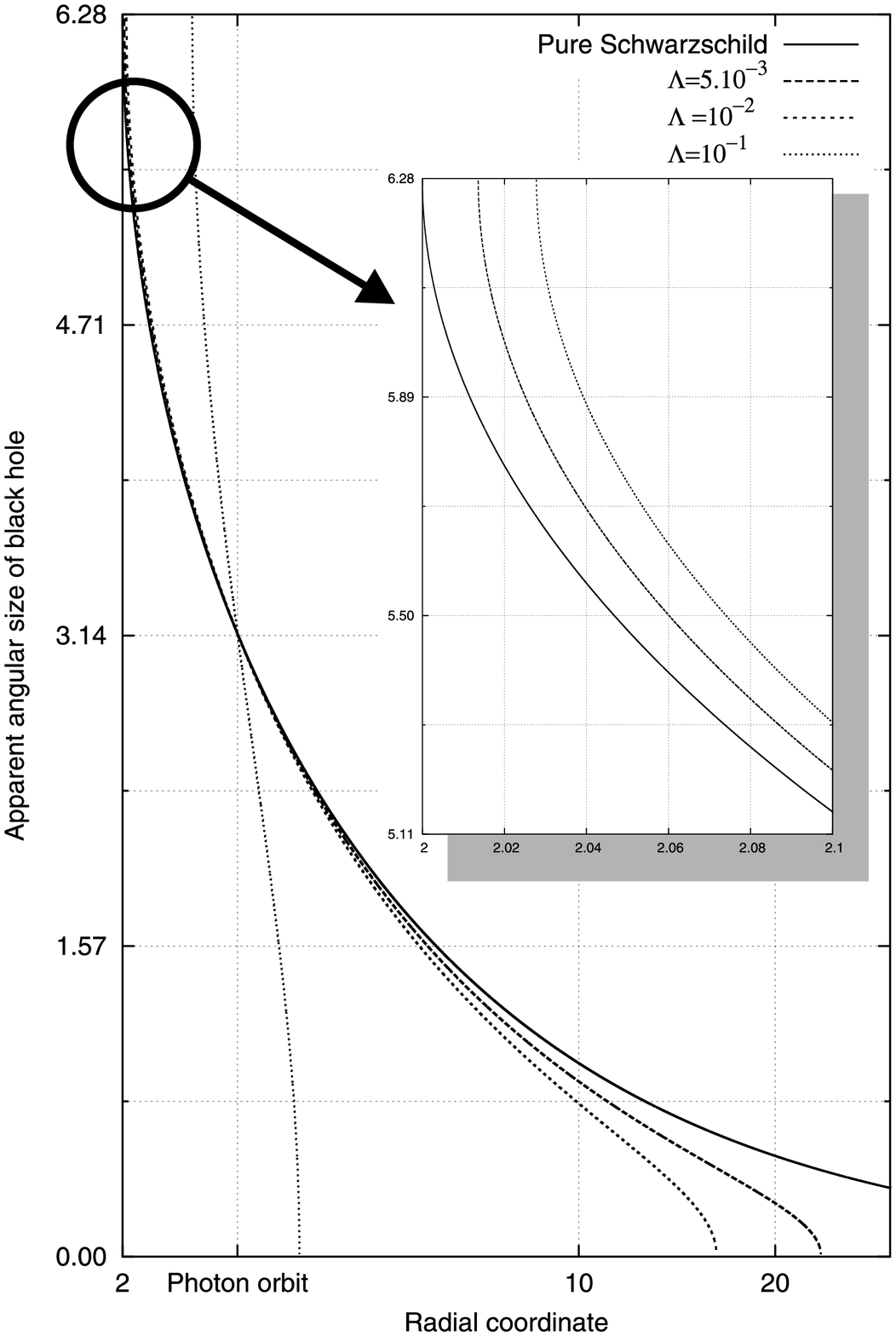}
\end{minipage}\hfill%
\begin{minipage}[b]{.48\hsize}
\centering
\includegraphics[width=\hsize]{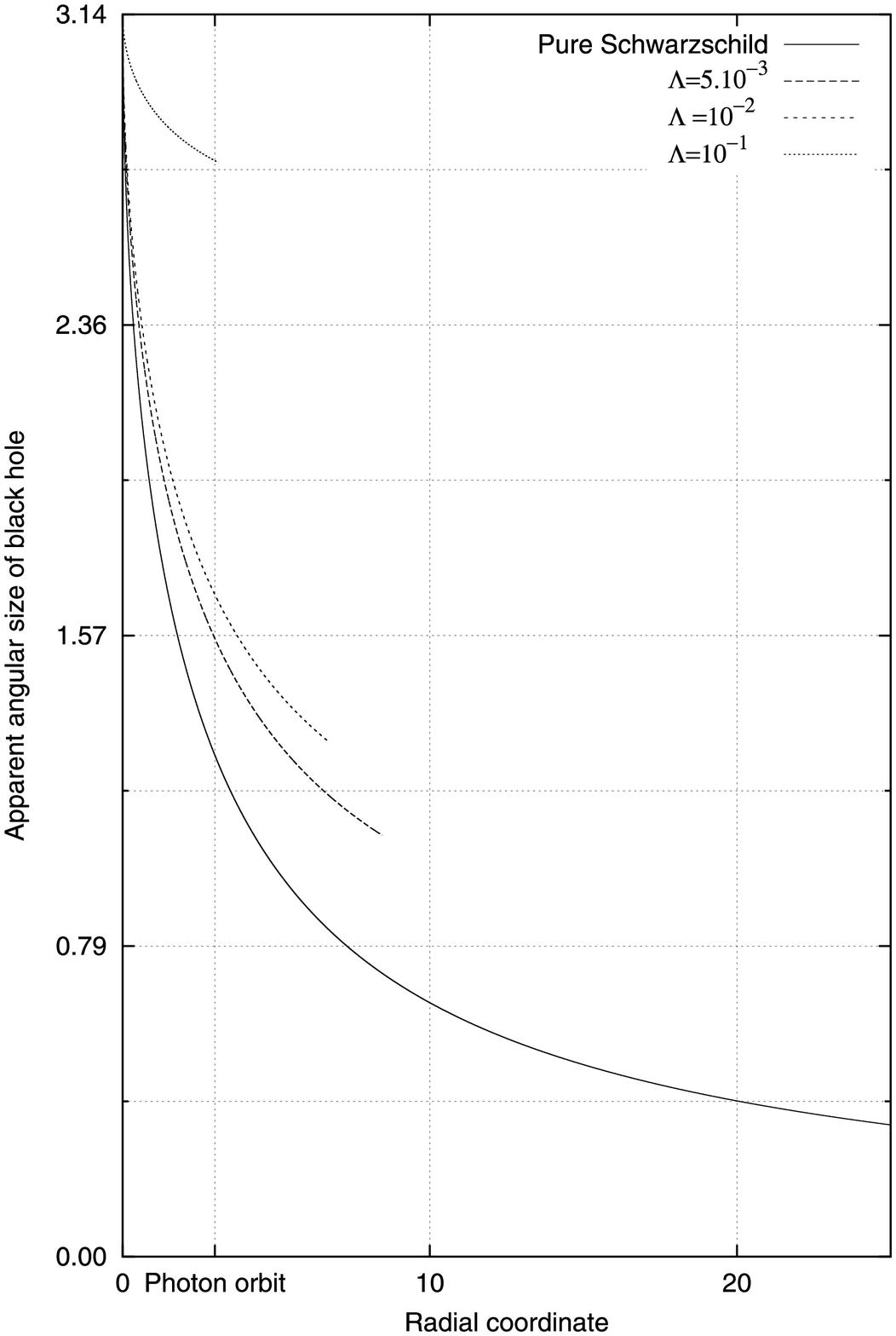}
\end{minipage}
\noindent
\caption{Apparent angular size of the black hole as function of observer's radial coordinate.
\textbf{Left panel:} static observers. \textbf{Right panel:} radially free-falling observers.}

\label{F1}
\end{figure}

The apparent angular size $\asize$ of the black hole can be naturally defined as the observed angular size of the circular black region on the observer sky, in which no images of distant objects can exist, and only radiation originated under the circular photon orbit can be observed \cite{BCHST,SP,cunn01}.
For observers located above the circular photon orbit the boundary of the black region corresponds to outgoing geodesics with $b$ approaching $b_{c}$ from above, while for observers located under the circular photon orbit the boundary corresponds to ingoing geodesics with $b$ aproaching $b_{c}$ from below.

In the case of static observers we have \cite{BCHST}
\begin{equation}
\asize = 2\arccos A(r_{\mathrm{obs}},y;b) \qquad  \mbox{,}\qquad  A(r,y;b)\equiv \pm\sqrt{1-\frac{b^{2}}{r^{2}}\left(1-\frac{2}{r}-yr^2\right) }\,.
\label{asizestat}
\end{equation}
Here $+\,\left(-\right)$ sign corresponds to observers located above (under) the circular photon orbit.
Above the circular photon orbit increasing $\Lambda$ causes downsizing of the black region, whereas under the circular photon orbit the black region grows with increasing $\Lambda$. In the limit case of observers located just on the circular photon orbit, $\asize$ is independent of $\Lambda$. It is invariably $\pi$, i.e., the black region always occupies just one half of the observer sky.

In the case of observers radially free-falling from the static radius \cite{BCHST,SH,SP}, the  apparent angular size of the black hole reads \cite{BCHST}

\begin{equation}\asize = 2\arccos  \frac{\left (Z(r_{\mathrm{obs}},y)+\sqrt{1-3y^{1/3}}A(r_{\mathrm{obs}},y;b) \right )} {\left( \sqrt{1-3y^{1/3}}+Z(r_{\mathrm{obs}},y)A(r_{\mathrm{obs}},y;b) \right )},
\label{asizefall}
\end{equation}
where
\begin{equation}
Z(r,y) \equiv \sqrt{\frac{2}{r}+yr^2-3y^{1/3}}.
\end{equation}
The $\Lambda$ dependency is qualitatively different. For radially free-falling observers $\asize$ grows with increasing cosmological constant at all values of the radial coordinate except the central singularity, where $\asize$ is invariably $\pi$, similarly to the case of static observers located on the circular photon orbit. Consequently, the radially free-falling observer will always observe smaller $\asize$ then the static observer at the same radial coordinate.

\section{Conclusions}

In this paper we discussed the influence of $\Lambda > 0$ on the optical projection in  strong, spherically symmetric gravitational field.
The influence depends on the value of the dimensionless cosmological parameter $y$. In the present universe with $\Lambda \sim {10}^{-56}\,\mathrm{cm}^{-2}$ values of $y$ are $y \sim 10^{-40}$ for stellar black holes and $y \sim 10^{-25}$ for supermassive black holes in  galactic nuclei. Observable effects can be expected for $y \ge 10^{-15}$ which corresponds to supergiant black holes with masses $M \ge 10^{15}\,\mathrm{M_{\odot}}$ \cite{SSH}.
In the case of  primordial black holes in the
very early universe, with assumed high values of repulsive cosmological
constant, one can expect even stronger effects. Considering
the electroweak phase transition at $T_{ew} \sim 100\,\mathrm{GeV}$,
we obtain an estimate of the primordial effective cosmological
constant $\Lambda_{ew} \sim  0.028\,\,\mathrm{cm}^{-2}$, while
considering the quark confinement at $T_{qc} \sim 1\,\mathrm{GeV}$ we obtain
$\Lambda_{qc} \sim 2.8 \times 10^{-10}\,\mathrm{cm}^{-2}$  and consequently higher values of $y$\,\cite{SSH}.

BHC code generates numerical solutions of the governing equation of the projection and static as well dynamic visualization outputs. Results show peculiar influence of $\Lambda$ on the apparent angular size of the black hole for observers in different local frames. This influence vanishes for static observers located at the circular photon orbit. For future studies we plan to extend our method and the BHC code in order to study axially symmetric spacetimes with repulsive cosmological constant.

\section*{Acknowledgement}
The present work was supported by the Czech grants MSM 4781305903 and LC06014 (P.\,B.). One author (P.\,B.)
would like to thank Eva \v Sr\' amkov\' a for useful discussions.

\section*{Appendix: tetrads and directly measured quantities}

It follows from the central symmetry of the geometry (\ref{metrika}) that
the geodetical motion of test particles and photons is allowed in the central
planes only. The existence of Killing vector fields $\xi_{(t)}$ and $\xi_{(\phi)}$ of the SdS
spacetime implies the existence of two constants of motion
\begin{equation}
  p_{\mathrm{t}}=g_{t\mu}p^{\mu}=-\mathcal{E},\qquad
  p_{\phi}=g_{\phi \mu}p^{\mu}=\Phi,
\end{equation}
and the photon motion is determined by the impact parameter
\begin{equation}
  b\equiv\frac{\Phi}{\mathcal{E}}.
\end{equation}
The 4-momentum of the photon reads \cite{SP}
\begin{equation}
  p_{\mathrm{t}} = -\mathcal{E},\qquad
  p_{\mathrm{r}} = \frac{A(r_{\mathrm{obs}},y,b)}{ B^{2}(r_{\mathrm{obs}},y)} \mathcal{E},\qquad
  p_{\phi} = b\mathcal{E} = \Phi,
\end{equation}
where  we introduce new variables
\begin{equation}
B^{2}(r,y) \equiv 1-\frac{2}{r}-yr^2\,,\qquad  A(r,y,b)=\pm \sqrt{1-B^{2}(r,y)\frac{b^2}{r^2}}.\label{A(r,y,b)}
\end{equation}
The + sign corresponds to photons receding from the black hole, while - sign  corresponds to photons infalling into the black hole.

In order to calculate directly measured quantities, one has to transform the 4-momentum of the photon into the local frame of the observer. The local components of 4-momentum for the observer at given $r_{\mathrm{obs}}$ can be obtained
using the appropriate tetrad of base 4-vectors $e_{\mu}^{(\alpha)}$, 1-forms $\omega_{\mu}^{(\alpha)}$ and transformation formulas
\begin{equation}
\omega^{(\alpha)}=e_{\mu}^{(\alpha)}dx^{\mu}\,,\qquad p^{(\alpha)}=e_{\mu}^{(\alpha)}p^{\mu}.
\end{equation}
\subsection*{Static observers}
The static observers located at rest at
$r=\mathrm{const}$, $\theta=\mathrm{const}$, $\phi=\mathrm{const}$
are endowed by a local frame with an orthonormal tetrad
of 1-forms \cite{SP}
\begin{equation}
  \omega^{(t)} = B(r,y)\,\mathrm{d} t\,,\qquad
  \omega^{(r)} = \frac{1}{B(r,y)}\,\mathrm{d} r\,,\qquad
  \omega^{(\theta)} =  r\, \mathrm{d} \theta\,,\qquad
  \omega^{(\phi)} = r \sin \theta \,\mathrm{d} \phi\,.
\end{equation}
The local components of \mbox{4-momentum} of the photon moving in the equatorial plane are given by the relations \cite{SP}
\begin{equation}
  p^{(t)}_{obs} =\frac{\mathcal{E}}{B(r_{\mathrm{obs}},y)}\,,\qquad
  p^{(r)}_{obs} = \frac{A(r_{\mathrm{obs}},y;b)}{ B(r_{\mathrm{obs}},y)} \mathcal{E}\,,\qquad
  p^{ (\phi)}_{obs} =\frac{ l \mathcal{E}}{r} =\frac{\Phi}{r_{\mathrm{obs}}}\,.
\end{equation}
Using general formulas (\ref{gen_Angle_shift}), the directional angle and the frequency shift are given as \cite{SP}
\begin{equation}
  \cos\alpha_{\mathrm{stat}} =  - A(r_{\mathrm{obs}},y,b)\,,\qquad  g_{\mathrm{stat}} =\frac{B(r_{\mathrm{source}},y)}{B(r_{\mathrm{obs}},y)}
\end{equation}
\subsection*{Observers radially free-falling from the static radius}
Local components and tetrads for free-falling observers can be obtained using Lorentz boost between the local frames of the static observer and a moving one at given $r_{\mathrm{obs}}$.
The orthonormal tetrad of 1-forms of appropriate local frame has
the form \cite{SP}
\begin{eqnarray}
  \omega^{(\tilde{t})}&=& \sqrt{1-3y^{1/3}}\, \mathrm{d} t
  + Z(r,y)B^{-2}(r,y)\, \mathrm{d}  r,\\
  \omega^{(\tilde{r})}&=& Z(r,y) \,\mathrm{d} t
  + \sqrt{1-3y^{1/3}}B^{-2}(r,y) \,\mathrm{d}  r,\\
  \omega^{(\tilde {\theta})}&=& r \,\mathrm{d} {\theta},\\
  \omega^{(\tilde {\phi})}&=& r  \sin{\theta}\,\mathrm{d} {\phi},
\end{eqnarray}
where  we introduced a new variable
\begin{equation}
Z(r,y) \equiv \sqrt{\frac{2}{r}+yr^2-3y^{1/3}}.
\end{equation}
The components of \mbox{4-momentum} of the photon measured by a observer radially free-falling from the static radius at a
given $r_{\mathrm{obs}}$  are given by the
relations \cite{SP}
\begin{eqnarray}
  p^{(\tilde{t})}_{\mathrm{obs}}&=& \frac{\mathcal{E}}{B^{2}(r_{\mathrm{obs}},y)}
  \left( \sqrt{1-3y^{1/3}}+Z(r_{\mathrm{obs}},y)A(r_{\mathrm{obs}},y;b) \right),\\
  p^{(\tilde{r})}_{\mathrm{obs}}&=& \frac{\mathcal{E}}{B^{2}(r_{\mathrm{obs}},y)}
  \left ( Z(r,y) + \sqrt{1-3y^{1/3}}A(r_{\mathrm{obs}},y;b) \right),\\
  p^{(\tilde{\phi})}_{\mathrm{obs}}&=& \frac{\mathcal{E} b}{r_{\mathrm{obs}}}
  = \frac{\Phi}{r_{\mathrm{obs}}}.
\end{eqnarray}
The directional angle and the frequency shift are given by the formulas \cite{SP}
\begin{equation}
\cos{\tilde{\alpha}_{\mathrm{fall}}}=
  - \frac{\left (Z(r_{\mathrm{obs}},y)+\sqrt{1-3y^{1/3}}A(r_{\mathrm{obs}},y;b) \right )}
  {\left( \sqrt{1-3y^{1/3}}+Z(r_{\mathrm{obs}},y)A(r_{\mathrm{obs}},y;b) \right )},\label{cospadaj}
  \end{equation}
\begin{equation}
  \tilde{g}_{\mathrm{fall}}\equiv \frac{p^{(\tilde{t})}_{\mathrm{obs}}}{p^{(t)}_{\mathrm{source}}}
  = \frac{B(r_{source},y)}{Z(r_{\mathrm{obs}},y)\cos{\tilde{\alpha}}
  +\sqrt{1-3y^{1/3}}}                                      \label{shift2}
\end{equation}

\vfill


\begin{thebibliography}{20}

\bibitem{BCHST} P. Bakala, P. \v Cerm\' ak, S. Hled\' ik, Z. Stuchl\' ik, K. Truparov\' a  Pl\v skov\' a: A virtual trip to the Schwarzschild-de Sitter black hole. Proceedings of RAGtime 6/7, 11-28, 2005

\bibitem{BEN1} W. Benger: Simulation of a Black Hole by Raytracing.  Relativity and Scientific Computing: Computer Algebra, Numerics, Visualization. Springer-Verlag Telos, 1996.

\bibitem{cunn01} C.T. Cunningham: Optical Appearance of Distant Observers near
  and inside a Schwarzschild Black Hole. Phys .Rev. D.12, 323--328, 1975.

\bibitem{ein01} A. Einstein: Lens-like action of a star by the deviation
  of light in the gravitational field.  Science 84, 506, 1936.

\bibitem{HAM1}  
A. J. S. Hamilton: Black Hole Flight Simulator. Bulletin of the American Astronomical Society, Vol. 36, p.810., 2004

\bibitem{KWR}
D. Kobras, D. Weiskopf, H. Ruder: Image-based rendering and general relativity. WSCG 2001 Conference Proceedings (pp. 130-137). University of West Bohemia, Pilsen, 2001.


\bibitem{Kra-Tur:1995:GenRelGrav:}
L.M. Krauss and M.S. Turner: The Cosmological constant is back.\\ Gen.Rel.Grav.27:1137-1144, 1995.

\bibitem{nem01} R.J. Nemiroff: Visual distortion near a neutron star a and
  black hole. Am.J.Phys. 61, 619, 1993.

\bibitem{NRHK} H.P. Nollert, H. Ruder, H. Herold and U. Kraus: The relativistic looks of a neutron star. Astron. Astrophys., 208, 153–156, 1989.

\bibitem{ohan01} H.C. Ohanian: The black hole as a gravitational
  ``lens''. Am. J. Phys. 55(5), 428--432, 1987.

\bibitem{Ost-Ste:1995:Nature:}
J.P. Ostriker, P.J. Steinhardt: The Observational case for a low density universe with a nonzero cosmological constant.
Nature 377:600-602, 1995.

\bibitem{our_web} Relativistic and particle physics and its astrophysical aplications,
Czech research project MSM 4781305903, http://www.physics.cz/research/.

\bibitem{SH} Z. Stuchl\' ik and. S. Hled\' ik: Some properties of the Schwarzschild--de~Sitter and Schwarzschild--anti--de~Sitter spacetimes.  Phys. Rev. D, 60(4):0044006(15 pages), 1999.

\bibitem{SP} Z. Stuchl\' ik and K. Pl\v skov\' a: Optical apperance of isotropically radiating sphere in the Schwarzschild--de~Sitter spacetime.  Proceedings of RAGtime4/5 , 167--185, 2004.

\bibitem{SSH} Z. Stuchl\' ik, P. Slan\' y and S. Hled\' ik: Equilibrium configurations of perfect fluid orbiting Schwarzschild–-de~Sitter black holes, Astron. Astrophys., 363, 425–-439, 2000. 

\bibitem{VIE1} S.U. Viergutz: Image generation in Kerr geometry. I. Analytical investigations on the stationary emitter-observer problem. Astron. Astrophys., 272, 355, 1993.


\end{thebibliography}
\end{document}